\documentclass[10pt,aps,pra,twocolumn,superscriptaddress]{revtex4-2}

\usepackage{bm} 
\usepackage{graphicx}
\usepackage{amsmath}    
\usepackage{latexsym}
\usepackage{amsfonts}   
\usepackage{amssymb}
\usepackage{comment}
\usepackage{array}      
\usepackage{txfonts}
\usepackage{xcolor}
\usepackage{float}
\usepackage{nameref}
\usepackage{placeins}
\usepackage{braket}
\usepackage{lipsum}

\usepackage[colorlinks=true,linkcolor=blue,urlcolor=blue,citecolor=blue,pdfusetitle]{hyperref}

\usepackage[caption=false]{subfig}

\usepackage{tikz}
\usepackage{tikz-3dplot}
\tdplotsetmaincoords{70}{110}
\usetikzlibrary{arrows.meta}
\usetikzlibrary{shapes.geometric, calc}

\begin{document}
\title{Long-Range Ergotropy Transfer with Weakened Boundary-Coupling Spin Chains}

\author{R. C. Krisnanda}
\affiliation{Department of Physics, Brawijaya University, Jl. Veteran, Malang 65145, Indonesia}

\author{M. S. Ukhtary}
\email{m.shoufie.ukhtary@brin.go.id}
\affiliation{Research Center for Quantum Physics, National Research and Innovation Agency (BRIN), South Tangerang 15314, Indonesia}

\author{Abdurrouf}
\email{abdurrouf@ub.ac.id}
\affiliation{Department of Physics, Brawijaya University, Jl. Veteran, Malang 65145, Indonesia}

\begin{abstract}
  Ergotropy is a key quantity of a quantum battery that measures the maximum extractable energy. When the extraction process is not local, a faithful transfer of ergotropy across distance is necessary. In this work, we investigate such transfer by using spin chains with weakened coupling at the boundaries as mediator. Weakening the coupling at the two boundaries of the chain gives rise to localized edge states allowing high fidelity transfer within given time duration. We found that, for long spin chains, the transferred ergotropy alternates between even and odd number of spins. We explain this parity effect by investigating the chain spectral structure. Furthermore, we consider both nearest-neighbor and long-range interaction between spin. In both model of interactions, weakening the boundary couplings improves the maximum transferred ergotropy for long spin chain, compared with in the case of conventional uniform chain. Although the long-range interaction reduces the transferred ergotropy, the transfer time is reduced, too, in particular for longer chain. In addition to the total transferred ergotropy, we also investigate the effect of initial states of the sender spin to the transport of the coherent and incoherent components of the total ergotropy, and the average transport efficiency.
\end{abstract}

\maketitle

\section{Introduction}

The continued miniaturization of electronic devices has increased interest in studying the quantum effect within the nanoscale electronic \cite{radamsonMiniaturizationCMOS2019,zengTransistorEngineeringBased2024,becher2023RoadmapMaterials2023,schuppSingleelectronDevicesSilicon2017}.
In the subsequent development of this technology, understanding how energy can be stored, manipulated, and extracted in quantum systems becomes increasingly relevant\cite{bhattacharjeeQuantumThermalMachines2021,kosloffQuantumHeatEngines2014a}.
One particular research direction is the study of quantum batteries, which are energy storage systems built around quantum mechanical effects.
Unlike conventional electrochemical batteries, which are commonly described by electrochemical processes, the quantum battery descriptions allow quantum mechanical properties to be used \cite{alickiEntanglementBoostExtractable2013a,campaioliColloquiumQuantumBatteries2024}.
Theoretical studies have shown that genuine properties of quantum mechanics, such as coherence and entanglement, can improve charging performance and storage capacity compared to a conventional battery \cite{campaioliEnhancingChargingPower2017,mohanReverseQuantumSpeed2021,crescenteChargingEnergyFluctuations2020,andolinaExtractableWorkRole2019}.
Experimental realizations of quantum batteries have also been demonstrated in various quantum system, including quantum dots \cite{maillettedebuywennigerExperimentalAnalysisEnergy2023}, superconductor \cite{huOptimalChargingSuperconducting2022}, organic microcavity \cite{quachSuperabsorptionOrganicMicrocavity2022a}, and qubit \cite{finkDressedCollectiveQubit2009}.

Energy storage and charging performance are some of the central questions of quantum-battery research \cite{ukhtaryHighperformanceKerrQuantum2023,downingChargingQuantumBattery2026,downingEnergyStorageContinuousvariable2025,downingEnergeticsPulsedQuantum2024,downingHyperbolicEnhancementQuantum2024,downingQuantumBatteryQuadratic2023b,downingTwophotonChargingQuantum2024,ferraroHighPowerCollectiveCharging2018,ahmadiSuperoptimalChargingQuantum2025,rodriguezCatalysisChargingQuantum2023,rodriguezOptimalQuantumControl2024,gangwarCoherentlyDrivenQuantum2024}.
The total energy stored in system, generally, is not equal to the amount of useful energy  that can be extracted from it.
The latter quantity, defined as ergotropy $\mathcal{E}$, measures the maximum extractable energy content of a quantum state under cyclic unitary operation \cite{allahverdyanMaximalWorkExtraction2004a,campaioliColloquiumQuantumBatteries2024}.
This subtle distinction is of importance subject, for it is possible a quantum state to possess non-zero stored energy while remaining having zero ergotropy.
In addition, when considering practical implementation of quantum battery, one encounters the problems of non-local ergotropy extraction when the battery and target subsystem are spatially separated.
This problem motivates us to investigate how the ergotropy can be transported.

\begin{figure}
  \centering

    \begin{tikzpicture}[x=1cm,y=0.8cm]
      \tikzset{point/.style={circle, fill=black, inner sep=2pt}}
      \usetikzlibrary{decorations.pathmorphing}
      \draw[] (0,0) -- (3,0);
      \draw[] (3,0) -- (3.5,0);
      \draw[dotted] (3.5,0) -- (4.5,0);
      \draw[]  (4.5,0) -- (5,0);
      \draw[] (5,0) -- (8,0);

      \node at (0.5, 0.5) {$\alpha J_1$};
      \node at (1.5, 0.5) {$J_1$};
      \node at (2.5, 0.5) {$J_1$};
      \node at (5.5, 0.5) {$J_1$};
      \node at (6.5, 0.5) {$J_1$};
      \node at (7.5, 0.5) {$\alpha J_1$};
      \node at (1, 1.7) {$\alpha J_2$};
      \node at (2, 1.7) {$J_2$};
      \node at (6, 1.7) {$J_2$};
      \node at (7, 1.7) {$\alpha J_2$};

      \node[circle,fill=white] at (0,0) {};
      \draw[thick,->,>=Stealth] (0,-0.25) -- (0,0.25);
      \foreach \x in {1,2,3,5,6,7,8}{
          \node[circle,fill=white] at (\x,0) {};
          \draw[thick,->,>=Stealth] (\x,0.25) -- (\x,-0.25);
        }

      \node[anchor=north] at (0, -0.4) {$1$};
      \node[anchor=north] at (1, -0.4) {$2$};
      \node[anchor=north] at (2, -0.4) {$3$};
      \node[anchor=north] at (6, -0.4) {$N-2$};
      \node[anchor=north] at (7, -0.4) {$N-1$};
      \node[anchor=north] at (8, -0.4) {$N$};

      \draw[dash pattern=on 6pt off 1pt] (0,0.6) .. controls (0,1.4) and (2,1.4) .. (2,0.6);
      \draw[dash pattern=on 6pt off 1pt] (1,0.6) .. controls (1,1.4) and (3,1.4) .. (3,0.6);
      \draw[dash pattern=on 6pt off 1pt] (5,0.6) .. controls (5,1.4) and (7,1.4) .. (7,0.6);
      \draw[dash pattern=on 6pt off 1pt] (6,0.6) .. controls (6,1.4) and (8,1.4) .. (8,0.6);
    \end{tikzpicture}

  \caption{The weakened coupling spin chain where the first site is modeled as qubit quantum battery, through which ergotropy is initially encoded. The ergotropy is transported to spin-$N$. The configuration consists of two sites, $1$ and $N$, that are weakly coupled to the rest of the spins by factor $\alpha<1$.} 
  \label{fig:chain}
\end{figure}
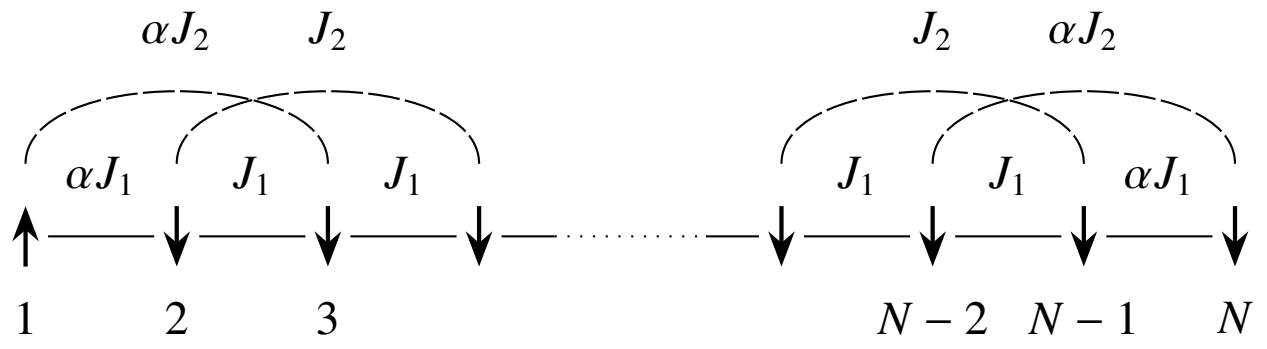

One possible approach to ergotropy transport is to transfer the quantum state of a battery, in which ergotropy is encoded, between spatially separated locations.
Quantum-state transfer has been extensively investigated in spin chains, where interactions between neighboring or more distant spins can mediate the propagation of quantum state along a one-dimensional system \cite{gualdiPerfectStateTransfer2008,christandlPerfectTransferArbitrary2005, boseQuantumCommunicationUnmodulated2003, wojcikUnmodulatedSpinChains2005,feldmanNearestneighborApproximationOneexcitation2023,moraguesAnisotropicHeisenbergSuSchriefferHeeger2026,petrosyanStateTransferStatic2010,burgarthConclusiveArbitrarilyPerfect2005,sousaPrettyGoodState2014,zwickSpinChainsRobust2012}.
Beyond theoretical consideration, experimental  realization of spin chains has ben realized in various platform \cite{senkoRealizationQuantumIntegerSpin2015,haukeCompleteDevilsStaircase2010,fengContinuousSymmetryBreaking2023}.
Perfect state transfer (PST), in which the received state is an exact replica of the initial state, provides a means of transporting ergotropy rather than merely transferring stored energy.
Assuming that the sender and the receiver have equivalent local Hamiltonian, this implies ergotropy of the transferred state is always preserved.

However, a simple realization of uniform nearest-neighbor spin chains, does not generally support PST for arbitrary long chain lengths \cite{boseQuantumCommunicationUnmodulated2003}.
In a well-known construction, the inter-spin couplings are engineered in such way to produce a spectrum and eigenstate structure whose contributions to the transfer amplitude interfere constructively at a specific time, thereby allowing PST \cite{christandlPerfectTransferArbitrary2005}.
Such engineered coupling, however, require precise fine-tuning, which for a long chain becomes increasingly impractical.
An alternative approach is to weaken the boundary couplings $\alpha J$ relative to those in the bulk $J$, where $\alpha$ controls the boundary-coupling strength as illustrated in Fig. \ref{fig:chain}  \cite{wojcikUnmodulatedSpinChains2005}.
Nevertheless, the transfer performance at PST regime may require long evolution times, particularly in regimes of weak boundary coupling $\alpha \to 0$. 
By incorporating the effect of long-range interaction, the transfer time can be generally be reduced at the cost of transfer performance \cite{gualdiPerfectStateTransfer2008}.
Regardless, such prolonged evolution can increase exposure to environmental effect such as dissipation, which may degrade the transferred state.
Hence, it is important to characterize not only the maximum transfer performance, but also attainable performance within relevant practical timescale.

Recent work by Murphy et al. \cite{murphyErgotropyTransportOnedimensional2025} has established the role of coherence in ergotropy transport through one-dimensional spin chains.
However, the thermodynamic consequences of the imperfect transfer, in particular the relationship between the evolution of receiver state with the contributions of population and coherence ergotropy, requires further investigation.
In addition, an even-odd effect in state transfer through weakened coupling spin chain has been observed previously \cite{wojcikUnmodulatedSpinChains2005}.
Nonetheless, the spectral structure and physical mechanisms underlying the parity dependence in the present setting, and their implications for the ergotropy transport within a finite observation time, warrant further investigation.
In this work, we address these question by characterizing the transfer of arbitrary qubit states using the weakened coupling spin chain as illustrated in Fig. \ref{fig:chain}, and by examining the evolution of the receiver's stored energy and ergotropy. 
We continue our investigation in the chain spectral structure to explain the previously observed even-odd behavior, in particular regarding the ergotropy transport. 
We extend the analysis to long-range interactions to examine how the interaction range affects the ergotropy transfer performance.

\section{method}

\subsection{Model of the spin chain}
We consider a finite one dimensional chain  of $N$ linearly arranged spin-$\tfrac12$ as illustrated in Fig. \ref{fig:chain}, each under the influence of uniform non-modulating field $B$.
Let us denote each spin site by $i$. We introduce a weakened coupling at boundary such that the system Hamiltonian is given by 
\begin{equation}
  H = -\sum_{i<j} J_{ij} \left( \sigma_x^i \sigma_x^{j} 
  + \sigma_y^i \sigma_y^{j} \right) - B\sum_{i=1}^{N}\sigma_z^i ,
  \label{eq:Hxx}
\end{equation}
where $\sigma_{x/y/z}^{i}$ denotes Pauli matrices acting on the $i$-th spin, and $J_{ij} = g_{i,j}/(a|i-j|)^\nu$ denotes the interaction between spins, where $a$ is the lattice distance and $\nu>0$ control its spatial decay.
We introduce dimensionless parameter $\alpha$ through
\begin{equation}
  g_{i,j}(\alpha)= 
  \begin{cases} 
    \alpha, & i=1\ \text{or}\ j=N, \\[2mm] 
    1, &\text{otherwise},
  \end{cases}
\end{equation}
to control the interaction between bulk and boundary spin.
Physically, this modification can be represented by an effective distance \(r_{\mathrm{eff}}=\alpha^{-1/\nu}(a|i-j|)\).
This formulation allow us to investigate the effect of long-range (LR) spin interaction and nearest-neighborhood (NN) approximation, where the latter is obtained by keeping only $|i-j|=1$ terms.
The case of $\alpha=1$ recovers the uniform coupling case, which will serve as our transfer benchmark, whereas the small $\alpha\ll1$ define weakened boundary coupling.
The XX Hamiltonian conserve excitation number $[H,\textstyle\sum_{i=1}^{N}\sigma_z^i]=0$, consequently the state evolution remain in one-excitation sector if the initial state also lies in the one-excitation sector.
We denote  $\ket{\mathbf{j}}$ as the one-excitation basis state in which the excitation is localized at site $j$.
Within this sector, the eigenvalue of NN regime is given by $E = -2B - 4J \cos k$, where $k$ must satisfy the transcendental equation \cite{wojcikUnmodulatedSpinChains2005}
\begin{equation}
  \eta\cot k \left[ \cot\left(\frac{N-1}{2}k\right) \right]^\eta  
  = \frac{\alpha^2}{2-\alpha^2},
  \label{eq:ktranscendental}
\end{equation}
with parity $\eta=\pm1$.
In uniform coupling $\alpha=1$, the spectrum  reduce into familiar $k = m\pi/(N+1)$, with integer $m=1,\ldots,N$.
As $\alpha$ introduced, two eigenstates become localized near the chain edges.
Assuming that the dominant eigenstates are these two edge-localized eigenstates, we have the following approximation eigenstate splitting 
\begin{equation}
  \epsilon_\mathrm{edge}
  =  \frac{4\alpha}{[a(N-1)]^\nu}.\label{eq:lrsplit}
\end{equation}
In the case $\alpha=0$, the boundaries become completely decoupled from the bulk and each other.
Since both sites experience the same uniform non-modulating $B$, their excitation energies are completely identical and zero splitting.
Consequently, degeneracy emerges both in full Hamiltonian and its boundaries restricted subspace.
Restoring LR interaction of both site via non-zero $\alpha \neq 0$ lift the degeneracy through their mutual interaction.
The resulting energy splitting increases with boundaries coupling strength $\alpha$, shortening the chain (decreasing $N$) or weakening the interaction spatial decay (decreasing $\nu$).

We consider the ergotropy transfer from the first site to $N$-th site. Therefore, the first site initially has excitation in it and the rest of the chain is set to be ground state $\ket{0}$ such that the state of the whole chain can be decomposed into $\rho(0) = \rho^{(1)}(0)\otimes_{i=2}^{N}\ket{0}\bra{0}$, where $\rho^{(1)}(0)$ is the initial density operator for the first spin.
We then turn on the interaction at $t=0$, and the state evolves as $\rho(t) = e^{-iHt}\rho(0)e^{iHt}$ under Hamiltonian given in Eq. \eqref{eq:Hxx} (we set $\hbar =1$). 
The state for $N$-th spin is obtained by taking partial trace of all spin state, except $N$-th spin. The density operator for the $N$-th spin is expressed in terms of initial state as follows \cite{murphyErgotropyTransportOnedimensional2025}:
\begin{equation}
  \rho^{(N)}(t)
  = \begin{bmatrix}
    \rho_{00}^{(1)}(0) + \rho_{11}^{(1)} (0)[1-|f_N|^2(t)] 
    & \rho_{01}^{(1)}(0)  f_N^*(t) \\
    \rho_{10}^{(1)}(0)  f_N(t)                  
    & \rho_{11}^{(1)}(0)|f_N|^2(t)
  \end{bmatrix},
  \label{eq:rhoN}
\end{equation}
 where $\rho_{ij}^{(1)}$ denotes the matrix element of the density operator $\rho^{(1)}$ and $f_N(t) = \braket{\mathbf{N } | e^{-iHt}|\mathbf{1}}$ describes the probability amplitude for an excitation initially at site 1 to arrive at site $N$ after time $t$.
In particular, $|f_N(t)| = 1$ corresponds to PST.
An important point is that the excited population of the $N$-th spin scales with $|f_N|^2$, whereas the coherence scales with $f_N$.
This scaling difference becomes important in our consideration to population and coherence associated transfer. The transfer amplitude can be expressed in terms of the eigenstates of $H$ as follows:
\begin{equation}
  f_N(t) = \sum_{j=1}^{N} e^{-i E_j t}\braket{\mathbf{N}|\tilde{m}_j }
  \braket{\tilde{m}_j |\mathbf{1} }.
  \label{eq:transamp}
\end{equation}
Under idealized condition of perfectly localized eigenstate at boundaries $\ket{\tilde {m}_\pm} = (\ket{\mathbf{1} }\pm \ket{\mathbf{N}})/\sqrt{2}$, the transfer dynamics is dominated by the symmetric and antisymmetric eigenstate pairs.
Transfer amplitude in Eq. (\ref{eq:transamp}) now reads $|f_N| = \sin (\Delta E t)/2$, and consequently PST occurs at $t=\pi/\Delta E$.
However, such localized state produce nearly degenerate eigenvalue, resulting in small  energy splitting, and consequently longer transfer time.

\subsection{Ergotropy}
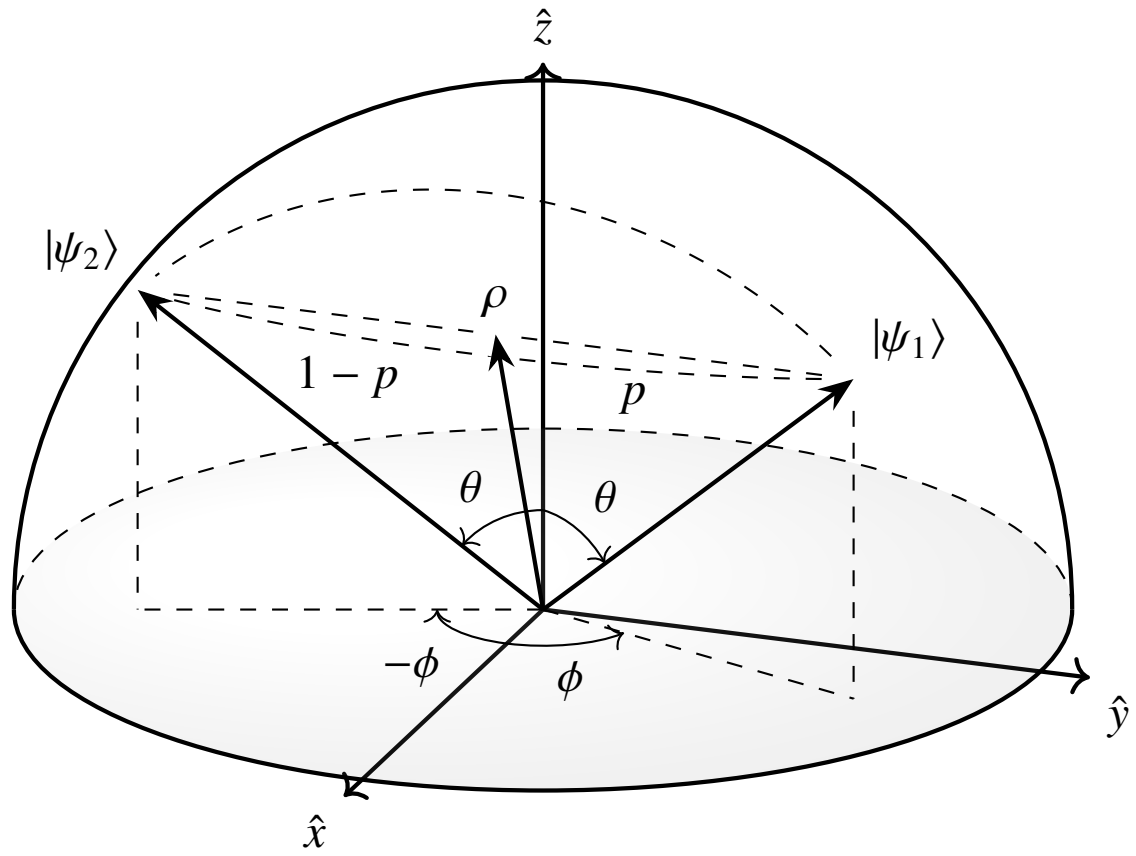
\begin{figure}
  \centering
  \begin{tikzpicture}[tdplot_main_coords,scale=3.5]
      \coordinate (O) at (0,0,0);
      \def\r{1}
      \def\th{50}
      \def\ph{70}
      \def\phm{-\ph}
      \pgfmathsetmacro{\rxy }{sin(\th)}

      \draw[thick,->] (O) -- (1.1,0,0) node[anchor=north east]{$\hat{x}$};
      \draw[thick,->] (O) -- (0,1.1,0) node[anchor=north west]{$\hat{y}$};
      \draw[thick,->] (O) -- (0,0,1.1) node[anchor=south]{$\hat{z}$};
      \shade[ball color=white, opacity=0.1] (0,0,0) circle (1);

      \tdplotsetcoord{A}{\r}{\th}{\ph}
      \tdplotsetcoord{Ath}{\r}{\th/2}{\ph}
      \tdplotsetcoord{A2}{\r}{\th}{\phm}
      \tdplotsetcoord{A2th}{\r}{\th/2}{\phm}
      \coordinate (P) at ({sin(\th)*cos(\ph)},0,{cos(\th)});

      \draw[dashed] (A) -- node[midway,midway,xshift=-15pt,yshift=-8pt, left] {$1-p$}  node[midway,xshift=20pt,yshift=-12pt, right] {$ p$} (A2) ;

      \tdplotdrawarc[->]{(O)}{0.2}{0}{\ph}{anchor=north}{$\phi$}
      \tdplotdrawarc[->]{(O)}{0.2}{0}{\phm}{anchor=north east}{$-\phi$}
      \tdplotsetthetaplanecoords{\ph}
      \tdplotdrawarc[->,tdplot_rotated_coords]{(O)}{0.2}{0}{\th}{anchor=south west
      }{$\theta$}
      \tdplotsetthetaplanecoords{\phm}
      \tdplotdrawarc[->,tdplot_rotated_coords]{(O)}{0.2}{0}{\th}{anchor=south east}{$\theta$}

      \draw[dashed]  (O) -- (Axy);
      \draw[dashed]  (A) -- (Axy);
      \draw[dashed]  (O) -- (A2xy);
      \draw[dashed]  (A2) -- (A2xy);

      \begin{scope}[tdplot_main_coords]
        \pgfmathsetmacro{\gamma}{acos(sin(\th)*sin(\th)*cos(\ph-\phm)+cos(\th)*cos(\th))}
        \draw[dashed,domain=0:1,samples=100,smooth]
        plot (
        {sin((1-\x)*\gamma)/sin(\gamma)*sin(\th)*cos(\ph)+sin(\x*\gamma)/sin(\gamma)*sin(\th)*cos(\phm)},
        {sin((1-\x)*\gamma)/sin(\gamma)*sin(\th)*sin(\ph)+sin(\x*\gamma)/sin(\gamma)*sin(\th)*sin(\phm)},
        {cos(\th)}
        );
        \draw[dashed,domain=0:1,samples=100,smooth]
        plot (
        {{sin(\th)*cos(\ph)}},
        {sin((1-\x)*\gamma)/sin(\gamma)*sin(\th)*sin(\ph)+sin(\x*\gamma)/sin(\gamma)*sin(\th)*sin(\phm)},
        {sin((1-\x)*\gamma)/sin(\gamma)*cos(\th)+sin(\x*\gamma)/sin(\gamma)*cos(\th)}
        );
      \end{scope}

      \node[circle,fill=white] at (A) {};
      \node[circle,fill=white] at (A2) {};
      \node[circle,fill=white] at (P) {};
      \draw[thick,->,>=Stealth] (O) -- (A) node [anchor=south west]{$\ket{\psi_1}$};
      \draw[thick,->,>=Stealth] (O) -- (A2) node [anchor=south east]{$\ket{\psi_2}$};
      \draw[thick,->,>=Stealth] (O) -- (P) node [anchor=south ]{$\rho$};

      \def\alp{20}
      \def\bet{70}
      \tdplotsetrotatedcoords{\alp}{0}{0}
      \begin{scope}[tdplot_rotated_coords]
        \draw[thick] (0,-\r) arc (-90:90:\r);
        \draw[dashed] (0,\r) arc (90:270:\r);
      \end{scope}

      \tdplotsetrotatedcoords{\alp}{\bet}{0}
      \begin{scope}[tdplot_rotated_coords]
        \draw[thick] (0,\r) arc (90:270:\r);
      \end{scope}
    \end{tikzpicture}

  \caption{Bloch sphere representation of the parameterized initial states of the first spin. The mixed state is constructed from a statistical mixture of two pure states $\ket{\psi_1}$ and $\ket{\psi_2}$, with probability $p$ and $1-p$, respectively}\label{fig:initialstate}
\end{figure}

After obtaining $\rho^{(N)}(t)$ given by Eq. \eqref{eq:rhoN}, the ergotropy can be calculated. The ergotropy is defined as the maximum extractable work from a quantum state $\rho$ through unitary operations, mathematically expressed as
\begin{equation}
  \mathcal E[\rho]=\operatorname{Tr}(\rho H)-\operatorname{Tr}(\rho_p H),\label{eq:ergo}
\end{equation}
where the first and second terms correspond to the stored energy and the passive energy, respectively. The passive energy is determined by so-called passive state $\rho_p$, from which no work can be extracted. Therefore, Eq. \eqref{eq:ergo} implies that not all energy stored in a QB might be useful for work. To calculate the passive energy, let us consider the spectral expansion of a state $\rho$ and a Hamiltonian $H$ as follows:
\begin{align}
  \rho & = \textstyle \sum_{j} r_j \ket{ r_j}\bra{ r_j}, \\  
  H & = \textstyle\sum_{k} \epsilon_k \ket{\epsilon_k}\bra{\epsilon_k},
\end{align}
where the eigenvalues are ordered as $r_1 \geq r_2 \geq \ldots$ and $ \epsilon_1 \leq \epsilon_2 \leq \ldots$, and the eigenstates of $\rho$ and $H$ are, respectively, denoted by $\ket{r_j}$ and $\ket{\epsilon_k}$.
Arbitrary unitary operations preserve eigenvalue spectrum of a quantum state, but allows its eigenvalues to be redistributed.
In particular, there exists unitary operation that distributes the largest eigenvalue to the lowest energy level, second-largest eigenvalue to second-lowest energy level, and so on. This unitary operation results the passive state, which maximizes the extractable work.  The resulting passive state is expressed as  \cite{allahverdyanMaximalWorkExtraction2004a}
\begin{equation}
  \rho_p = \sum_{j}r_j \ket{\epsilon_j} \bra{\epsilon_j},
\end{equation}
where the passive energy is derived as $E_p\equiv \operatorname{Tr}(\rho_p H)=\sum_{j}r_j\epsilon_j$.

For a general qubit state with Hamiltonian $H = -B\sigma_z$, this reads
\begin{equation}
  \mathcal E[\rho]=B\left(2\rho_{11}-1+\sqrt{1-4\det\rho}\right).
  \label{eq:ergotot_gen}
\end{equation}
Ergotropy in qubit originates via two mechanisms, either population inversion and/or coherence in energy eigenbasis \cite{francica2020quantum}.
The former contribution become non-zero if the higher energy level is more populated, i.e. $\rho_{11}>1/2$, and can be conveniently expressed as 
\begin{equation}
  \mathcal E_{\mathrm{pop}}[\rho]=2B\max(2\rho_{11}-1,0).
  \label{eq:ergopop_gen}
\end{equation}
The latter contribution accounts for ergotropy associated with the off-diagonal elements, simply defined as $\mathcal E_{\mathrm{coh}}(\rho)=\mathcal E(\rho)-\mathcal E_{\mathrm{pop}}(\rho)$.
Explicitly,
\begin{equation}
  \mathcal E_{\mathrm{coh}}[\rho]=B\left[\sqrt{(1-2\rho_{11})^2+4|\rho_{01}|^2}-|1-2\rho_{11}|\right].
  \label{eq:ergocoh_gen}
\end{equation}
Alternatively, we can express ergotropy using the Bloch sphere representation by writing the state as $\rho = (\mathbf 1+\mathbf r\cdot\boldsymbol\sigma)/2$, where $\mathbf {r}\equiv (x,y,z)$ is the Bloch vector.
The ergotropy becomes
\begin{equation}
  \mathcal E[\rho]=B(|\mathbf r|-z)=B|\mathbf r|(1-\cos\theta),
  \label{eq:ergo_bloch}
\end{equation}
with $\theta\equiv z/|\mathbf r|$. Likewise, the population and coherence contribution, respectively, can be written as 
\begin{align}
  \mathcal E_{\mathrm{pop}} [\rho] & = 2B\max(-z,0),\\
  \mathcal E_{\mathrm{coh}} [\rho] & = B\left(|\mathbf r|-|z|\right).
\end{align}
These expressions provide the geometric interpretation of the ergotropy, where $\mathcal E_{\mathrm{pop}}$ measures the negative-\(z\) component of the vector $\mathbf{r}$, whereas $\mathcal E_{\mathrm{coh}}$ determines the contribution associated with the vector's transverse components.

In this work, we investigate how initial states influence the ergotropy transport.
Hence, to consider a general state of sender qubit, we express our initial state as statistical mixture of two pure states as follows:
\begin{equation}
  \rho^{(1)}(0)=p\ket{\psi_1}\bra{\psi_1}+(1-p)\ket{\psi_2}\bra{\psi_2},\label{eq:mixed}
\end{equation}
where $p$ is the state probabilities weight of the first pure state and $\ket{\psi_i}=\cos({\theta_i}/{2})\ket{0}+e^{i\phi_i}\sin({\theta_i}/{2})\ket{1}$ represents a general qubit pure state on the Bloch sphere.
To reduce the number of independent parameters, we impose the constraint of $\theta_2=\theta_1\equiv\theta$ and $\phi_2=-\phi_1\equiv-\phi$.
This parametrization allows us to describe a general qubit state in terms of two pure states related by reflection across $xz$-plane, as shown in Fig. (\ref{fig:initialstate}). Note that for general $p$, $\rho^{(1)}(0)$ is a mixed state, which is described by the Bloch vector inside the Bloch sphere as illustrated by Fig. (\ref{fig:initialstate}). 
Substituting these constraints yields
\begin{equation}
  \rho^{(1)} (0) =
  \begin{bmatrix}
    \cos^2\frac{\theta}{2}
    & \frac{\sin\theta}{2}\left(e^{i\phi}-i2p\sin\phi\right) \\[2mm]
    \frac{\sin\theta}{2}\left(e^{-i\phi}+i2p\sin\phi\right)
    & \sin^2\frac{\theta}{2}
  \end{bmatrix}.
\end{equation}
This sender qubit is then coupled with the remaining site, and the interaction is then  turned on. 
After time $t$, the receiver state according to Eq. (\ref{eq:rhoN}) is given by
\begin{equation}
  \rho^{(N)}(t)=
  \begin{bmatrix}
    1-\sin^2\frac{\theta}{2}|f_N|^2
    & \frac{\sin\theta}{2}\left(e^{i\phi}-i2p\sin\phi\right)f_N\\[2mm]
    \frac{\sin\theta}{2}\left(e^{-i\phi}+i2p\sin\phi\right)f_N^*
    & \sin^2\frac{\theta}{2}|f_N|^2
  \end{bmatrix}.
  \label{eq:rhoN_const}
\end{equation}
Having obtained the receiver density state Eq. (\ref{eq:rhoN_const}), we can now evaluate ergotropy delivered in receiver state.
Using general ergotropy expression in Eq. (\ref{eq:ergotot_gen}), we have
\begin{multline}
    \mathcal E[\rho^{(N)}]
=B\Bigg(
 2|f_N|^2\sin^2\frac{\theta}{2}-1\\
+\bigg\{1-4|f_N|^2\big[p(1-p)\sin^2\theta\sin^2\phi\\
 +\sin^4\frac{\theta}{2}(1-|f_N|^2)\big]\bigg\}^{1/2}
\Bigg).
\label{eq:ergotot_const}
\end{multline}
In addition, its population and coherence contributions, respectively, are given by (\ref{eq:ergopop_gen}) and (\ref{eq:ergocoh_gen}) read
\begin{align}
  \mathcal E_{\rm pop}(\rho^{(N)})
  & = 2B\max\left(2|f_N|^2\sin^2\frac{\theta}{2}-1,0\right)
  \label{eq:ergopop_const}\\
  \mathcal E_{\rm coh}(\rho^{(N)})
  & = B\Bigg(
    \bigg\{
      \left(1-2|f_N|^2\sin^2\frac{\theta}{2}\right)^2\notag\\
      &\qquad+|f_N|^2\sin^2\theta
      \left(\cos^2\phi+(1-2p)^2\sin^2\phi\right)
    \bigg\}^{1/2}\notag\\
    &\qquad-\left|1-2|f_N|^2\sin^2\frac{\theta}{2}\right|
    \bigg].
  \label{eq:ergocoh_const}
\end{align}
Besides the transfer amplitude $f_N$, the transferred ergotropy in Eq. (\ref{eq:ergotot_const}) together with its components in Eqs. (\ref{eq:ergopop_const}) and (\ref{eq:ergocoh_const}) also depend on the parameters of the initial mixed state defined in Eq. \eqref{eq:mixed}.
Consequently, different initial state may produce different ratio of the initially encoded ergotropy $\mathcal{E}[\rho^{(1)}(0)]$ and the  received ergotropy $\mathcal{E}[\rho^{(N)}(t)]$.
Characterization based only on initial state, therefore, insufficient to describe overall channel performance.
To address this problem, we characterize the transfer performance by averaging the state-dependent efficiency uniformly over all physically accessible initial qubit states.
For non-zero initial encoded ergotropy, we define the average performance efficiency as 
\begin{equation}
  \eta_N(t)
  =\frac{1}{V}\int_{\mathcal B}dV\;\frac{\mathcal E[\rho^{(N)}(t)]}{\mathcal E[\rho^{(1)}}\label{eq:eff}
\end{equation}
where $V=4\pi/3$ is the Bloch-ball volume and the integration is over the whole Bloch sphere.

\section{Results and discussions}

\begin{figure}
    \centering
    \includegraphics[width=\linewidth]{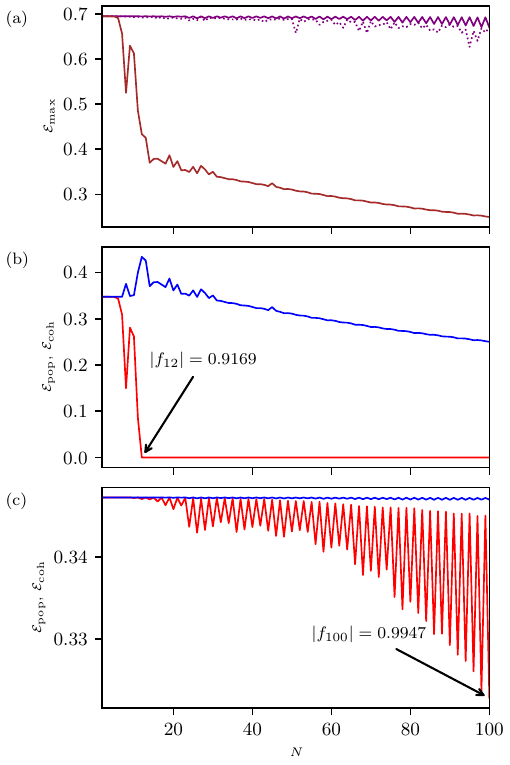}
    \caption{Maximum $\mathcal E[\rho^{(N)}(t)]$ over the time interval $t\in(0,1000)$ as a function of chain length $N$, for equal population and coherence ergotropy contribution $\mathcal E_{\mathrm{pop}}[\rho^{(N)}] = \mathcal E_{\mathrm{coh}}[\rho^{(N)}] = 0.3472$, specified by $p=0.2506$, $\theta=100^\circ$, $\phi=90^\circ$.  (a) The total $\mathcal E[\rho^{(N)}(t)]$ due to weakened coupling and uniform coupling are shown by purple and brown curve, respectively. Solid and dashed purple curve correspond to optimized and fixed $\alpha$ across $N$ respectively. (b) $\mathcal E_{\mathrm{pop}}[\rho^{(N)}]$ (red) and $\mathcal E_{\mathrm{coh}}[\rho^{(N)}]$ (blue) contributions for uniform and (c) weakened coupling.}
    \label{fig:ergocontribution}
\end{figure}

\begin{figure*}
    \centering
    \includegraphics[width=\textwidth]{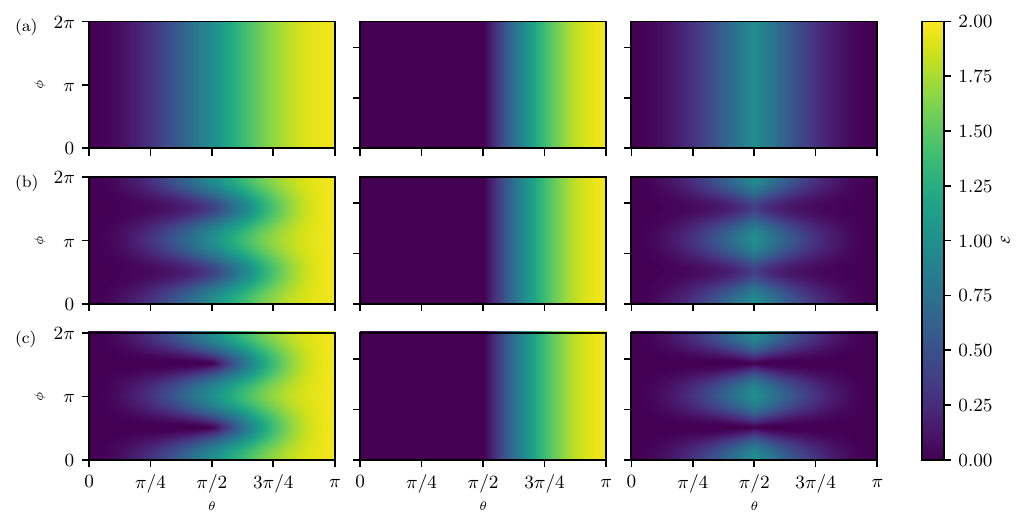}
    \caption{The transferred ergotropy heatmaps over the time interval $t\in(0,1000)$ as functions of the initial-state parameters $\theta$ and $\phi$ for $N=100$ and $\alpha = 0.0306$. From left to right, the columns show the $\mathcal E[\rho^{(N)}(t)]$, $\mathcal E_{\mathrm{pop}}[\rho^{(N)}]$ and $\mathcal E_{\mathrm{coh}}[\rho^{(N)}]$ contributions to $\mathcal E[\rho^{(N)}(t)]$], respectively. The rows correspond to (a) $p=0$, (b) $p=0.3$, and (c) $p=0.5$.}
    \label{fig:heatmap}
\end{figure*}

To investigate the transport of $\mathcal E_{\mathrm{pop}}[\rho]$ and $\mathcal E_{\mathrm{coh}}[\rho]$, we consider initial states having ergotropy with equal contributions from both origins, given by parameters $p=0.2506$, $\theta=100^\circ$, $\phi=90^\circ$. 
Figure \ref{fig:ergocontribution} (a) compares the maximum receiver ergotropy $\mathcal E[\rho^{(N)}(t)]$ for NN weakened and uniform couplings, represented by the purple and brown curves, respectively, as a function of $N$. 
The weakened-coupling configuration consistently outperforms the uniform chain, in particular for large \(N\), where the transferred ergotropy achieves almost the same value with the initial ergotropy. 
For each chain length, we optimize the boundary-coupling parameter \(\alpha\) within \(0<\alpha<1\) to maximize $\mathcal E[\rho^{(N)}(t)]$ within a fixed period of waiting time. 
As expected, this optimized choice (solid curve) yields better performance than using a fixed value of \(\alpha\) (dashed curve) across all chain lengths. 
Furthermore, the weakened-coupling results exhibit even-odd dependence on \(N\) for large \(N\), consistent with previously reported parity effects in state transfer through weakened-coupling spin chain. For large $N$, the maximum ergotropy of odd-chain with optimized $\alpha$ is larger that of even-chain. 
This parity effect can be understood by considering the spectral structure of NN Hamiltonian.
The spectrum of odd chain contains a band center eigenvalue of $E=-2B$ with eigenstate $\ket{m_0}$, which under weakened coupling $\alpha\ll1$ regime, is guaranteed to be localized at the chain boundaries $\ket{m_0} = (\ket{\mathbf{1}}-\ket{\mathbf{N}})/ {\sqrt{2}}$. 
In addition to other two near center bands, this band center state $\ket{m_0}$ contributes $1/2$ to the total $f_N(t)$.
For even chain, however, there is no band center eigenvalue. The two central states $\ket{m_\pm}\equiv (\ket{\mathbf{1}}\pm\ket{\mathbf{N}})/ {\sqrt{2}}$ contribute essentially all the endpoint transfer, therefore their localization must be relatively strong before the two-state approximation works well.
Consequently, substantial ergotropy transfer can persist in odd-chain, even when the neighboring modes are less strongly localized at the edges.
In addition, further explanation for this parity effect can be made from the energy splitting between the dominant states. 
Expanding Eq. \eqref{eq:ktranscendental} around band-center gives the energy splitting between dominant states
\begin{equation}
  \epsilon = \begin{cases} \dfrac{4\alpha^2}{2-\alpha^2}\simeq2\alpha^2,  & N\; \mathrm{even}, \\\\
              4\sqrt{\dfrac{2\alpha^2}{(N-1)(2-\alpha^2)}}\simeq \dfrac{4\alpha }{\sqrt{N-1}}, & N\ \mathrm{odd}
  \end{cases}.
\end{equation}
The different scaling with $\alpha$ means that, in the regime $\alpha\ll 1$, the odd-chain splitting is larger than the even-chain splitting. 
Since the characteristic transfer time scales inversely with the energy splitting $t \sim 1/\epsilon$, the odd-chain can reach substantial transferred ergotropy more quickly. Thus, for a finite waiting time period, the ergotropy of odd-chain is larger than that of even-chain.

Figure \ref{fig:ergocontribution} (b) and (c) show the population (red) and coherence (blue) ergotropy contribution to total ergotropy for uniform coupling and for weakened coupling, respectively.
As $N$ increases, the population contribution appears to decay more rapidly than the coherence contribution.
A notable difference is observed between two coupling configuration: the population contribution vanishes for chain beyond $N=12$ under the uniform coupling, whereas weakened coupling preserve the contribution throughout even for $N=100$.
This is behavior can be understood from transfer amplitude threshold imposed by Eq. \eqref{eq:ergopop_const}, specifically the population contribution vanishes if $|f_N| \leq 1/\sqrt{2\sin (\theta/2)}$.
For the chosen initial state with $\theta=100^\circ$, this corresponds to transfer amplitude smaller than $|f_N| < 0.923$.
While the weakened coupling satisfies this condition across the investigated chain lengths, uniform coupling fails to do so  for sufficiently long chain.
Interestingly, under uniform coupling near $N=10$, the coherence ergotropy contribution can exceed its initial value, despite imperfect state transfer, which indicate the contribution does not decrease monotonically with respect to transfer amplitude.
To further investigate how the imperfect transfer affects these contributions we seek to examine the ergotropy transport across different initial states.
Fig. \ref{fig:heatmap} show the distribution of total, population and coherence ergotropy.
The column from left depict total, population, and coherence ergotropy respectively, while, the rows show different value of parameters $p=(0,0.3,0.5)$.
Since the ergotropy dependence to parameter $p$ enters in the term of $p(1-p)$, the ergotropy is invariant under the transformation $p\to 1-p$.
Consequently, the range $0\leq p \leq 0.5$ is sufficient to represent ergotropy dependence with this parameter.
For $p=0$, corresponding to pure initial state, both ergotropy and its contribution are invariant with $\phi$, which reflects ergotropy invariance with azimuthal rotation as shown by Eq. \ref{eq:ergo_bloch}.
The initial state, then, become mixed as $p$ increase to $0.3$ and $0.5$.
In this regime, the ergotropy is no longer invariant with respect to $\phi$.
Increasing $p$ makes the ergotropy azimuthal dependence more pronounced instead.
We identify a region in which the initial state has nonzero population and coherence contributions, whereas the receiver, due to imperfect transfer, retains only a coherence contribution.
This family of state is given by $\pi/2<\theta<2 \arcsin(1/\sqrt{2|f_N|^2})$, with the upper boundary roughly $1.5815$ rad for $N=100$.
In contrast, we find no corresponding region in which only the population contribution remains receiver.
The physical mechanism behind this can be made clear by expressing the receiver Bloch vector $(\mathbf r^{(N)})$ in term of initial sender Bloch vector $(\mathbf r^{(1)})$ as follows \cite{palamaraMagicTransferQuantum2026}:
\begin{equation}
  \mathbf r^{(N)}  =  D(|f_N|)\,R_z(\delta) \mathbf r^{(1)} +\mathbf t(|f_N|)
\end{equation}
where 
\begin{gather}
  R_z (\delta) =
  \begin{bmatrix}
    \cos\delta  & -\sin\delta & 0 \\
    \sin\delta & \cos\delta & 0 \\
    0           & 0          & 1
  \end{bmatrix}\\
  D(|f_N|) = \operatorname{diag}(|f_N|,|f_N|,|f_N|^2)\\
  \textbf{t} (|f_N|)= \begin{bmatrix} 0, 0,1-|f_N|^2 \end{bmatrix}
\end{gather}
The complex transfer amplitude $f_N = |f_N|e^{i\delta}$ introduce rotation about $z$ axis by angle $\delta$, represented by $R_z(\delta)$.
This does not have any effect on ergotropy transfer since ergotropy is invariant to azimuthal rotation.
The diagonal matrix $D$ then produces anisotropic contraction where it acts differently on transverse $(x,y)$ and longitudinal $(z)$ component of $\mathbf r^{(1)}$.
Finally, $\textbf{t}$ add affine displacement along the \(+z\) direction which represent relaxation toward the ground-state. 
This contraction and displacement explain why imperfect transfer reduces population ergotropy more than coherence ergotropy.
Geometrically, for a state to contain both population and coherence ergotropy, it must lie somewhere on southern hemisphere of the Bloch sphere.
The solid red line, then, denote the state in Bloch sphere that gets mapped out into northern hemisphere by imperfect transfer.
In contrast, a receiver state with population ergotropy alone has to lie on the negative $z$ axis.
Such state, however, is not possible to obtain under imperfect transfer $f_N \neq 0$.
Furthermore, both contraction and affine translation can reduce the magnitude of its longitudinal component $|z|$ faster than its Bloch vector length $|\mathbf{r}|$ decreases.
Consequently the coherence ergotropy $\mathcal E_{\mathrm{coh}}=B(|\mathbf r|-|z|)$ can increase, even as the transverse components contract.

\begin{figure}
  \centering
  \includegraphics[width=\linewidth]{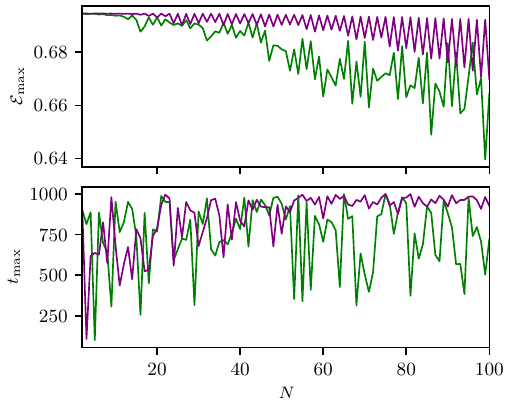}
  \caption{(a) Maximum $\mathcal E[\rho^{(N)}(t)]$ throughout the time interval $t\in(0,1000)$ with $\nu=3$ (long-range interaction) and (b) the corresponding transfer time $t_{\max}$ as functions of chain length $N$. The purple and green curves represent nearest-neighbor (NN) and long-range (LR) interactions, respectively.}
  \label{fig:lrinteraction}
\end{figure}

\begin{figure}
  \centering
  \includegraphics[width=\linewidth]{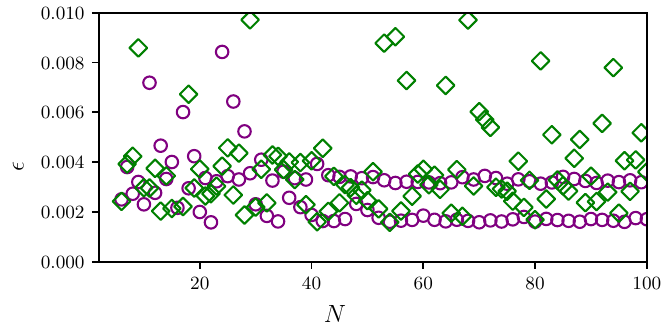}
  \caption{Numerical energy splitting $\epsilon$ across chain length $N$. Purple circles and green diamonds denote NN and LR interactions, respectively}
  \label{fig:lrenergy}
\end{figure}

\begin{figure}
  \centering
  \includegraphics[width=\linewidth]{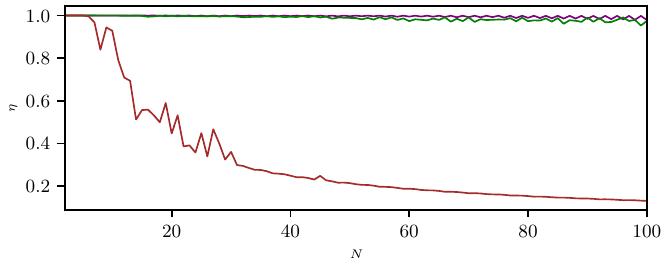}
  \caption{Figure: Average transfer efficiency as a function of chain length $N$. The purple and green curve brown curves denote NN and LR model with weakened coupling.  The uniform coupling, shown by brown curve, serve as benchmark. Solid and dashed line represent closed system and open system performance respectively}\label{fig:eff}
\end{figure}

We  continue our investigation on the effect of long-range (LR) interactions. In
Fig. \ref{fig:lrinteraction}, we show the maximum $\mathcal E[\rho^{(N)}(t)]$. along the chain with $\nu=3$ for the coupling strength [See Eq. \eqref{eq:Hxx}], corresponding to dipole-dipole interactions with couplings decaying as $r^{-3}$.
As in the case with NN, we also optimize $\alpha$ for each $N$. 
Compared with the NN case, the LR model generally exhibits lower maximum ergotropy but reaches its maximum at shorter time especially for large $N$. 
This behavior is accompanied by an increased energy splitting between the dominant eigenstates, as shown in Fig. \ref{fig:lrenergy}.
This trend is consistent with the energy splitting approximation given in Eq. \eqref{eq:lrsplit}, which states that the energy splitting is conversely proportional to characteristic interaction $\nu$. 
These results highlight a trade-off between transfer performance with transfer time, where the LR model of interaction can accelerate ergotropy transfer, but this speedup comes at the cost of reduced maximum transfer performance.
Furthermore, we compare the models using the average efficiency defined in Eq. \eqref{eq:eff}, with the results shown in Fig. \ref{fig:eff}.
NN model achieves the greatest average efficiency among the considered model, nonetheless weakened coupling still able to improve LR model compared to uniform coupling.

\section{Conclusion}

We have investigated the ergotropy transfer in spin chains with weakened boundary-couplings, considering both nearest-neighbor (NN) and long-range (LR) interactions. 
In both models of interaction, weakening the boundary couplings improves the maximum receiver ergotropy compared with the uniform coupling chain. 
The distinct spectral structures of NN chains with even and odd $N$ reflect to the ergotropy transfer performance. 
Although the LR model generally achieves lower maximum receiver ergotropy, it can reduce the transfer time, particularly for longer chains. 
These results highlight a trade-off between transfer performance with transfer time.

Moreover, we investigate the transfer of population and coherent contributions of the total ergotropy across initial states. We identify a region in which the initial state has nonzero population and coherence contributions, whereas the receiver, due to imperfect transfer, retains only a coherence contribution. On the other hand, the opposite case is not possible. We have explained this behaviour by using the transformation of the initial Bloch vector, which contracts the transverse components, while the longitudinal component is both contracted and displaced toward the ground-state pole.
This transformation explains why the coherent contribution is generally better preserved than population contribution.
Additionally, this also explains why coherent contribution may exceed its initial value.
Overall, these results establish the weakened boundary-couplings as a practical means of improving ergotropy transfer in long spin chains.
Moreover, understanding how imperfect state transfer affect the receiver ergotropy and its components provides insight into the design of quantum energy-storage applications.

%

\end{document}